# Rotating vortex solitons supported by localized gain

Olga V. Borovkova, Valery E. Lobanov, Yaroslav V. Kartashov, and Lluis Torner

*ICFO-Institut de Ciencies Fotoniques, and Universitat Politecnica de Catalunya, Mediterranean Technology Park, 08860 Castelldefels (Barcelona), Spain*
*Corresponding author: Olga.Borovkova@icfo.es

We show that ring-like localized gain landscapes imprinted in focusing cubic (Kerr) nonlinear media with strong two-photon absorption support new types of stable higher-order vortex solitons containing multiple phase singularities nested inside a single core. The phase singularities are found to rotate around the center of the gain landscape, with the rotation period being determined by the strength of the gain and the nonlinear absorption.

Vortex solitons are self-sustained excitations carrying a nonzero angular momentum and topological phase singularities [1]. In uniform focusing local media, vortex solitons exhibit ring-like profiles, a property that makes them prone to azimuthal modulation instabilities. Such instability can be suppressed in conservative materials by several mechanisms, including competing nonlinearities, external potentials, or nonlocalities. Stable vortex solitons form also in dissipative settings [2,3]. Dissipative vortex solitons have been found in laser amplifiers [4], in Bose-Einstein condensates [5,6], and in systems described by the complex cubic-quintic Ginzburg-Landau equation [7-9]. Vortex solitons in uniform dissipative media may experience considerable dynamical shape deformations, as it occurs in suitable Ginzburg-Landau systems [8,9] and in wide-aperture lasers with a saturable absorber [4].

Recently it was predicted that the evolution of nonlinear excitations in dissipative medium is affected dramatically by a spatially modulated gain. Such evolution has been studied in Bragg gratings and optical waveguides [10,11], in materials with periodic refractive index modulation [12,13], and in Bose-Einstein condensates [14]. In such systems solitons form due to the interplay between localized gain and nonlinear absorption, and a spatial localization of the gain landscape ensures the stability of the background field. Such interplay may result in suppression of azimuthal modulation instabilities and formation of stable stationary vortex solitons [15]. However, the possibility to excite rotating vortex states in such systems has never been addressed to date.

In this Letter we reveal that ring-like gain landscapes imprinted in cubic media with two-photon absorption support rich families of rotating vortex states exhibiting multiple phase singularities embedded into a single vortex core and possessing azimuthally modulated shapes. Notice that while in some conservative systems the formation of vortex solitons with multiple singularities may be stimulated by external factors, such as large sample asymmetry in nonlocal media [16] or the presence of strong optical lattices [17], in our dissipative setting the asymmetric rotating vortices appear despite the fact that all parameters are either spatially uniform (diffraction and nonlinearity) or radially symmetric (gain). This indicates that the formation of such states is mediated by complex internal energy flows that may be highly asymmetric in a system with localized gain.

We consider the propagation of laser radiation along the $\xi$-axis in a cubic nonlinear medium with strong two-photon absorption and nonuniform gain that is described by the nonlinear Schrödinger equation for the field amplitude $q$:

$$i\frac{\partial q}{\partial \xi} = -\frac{1}{2}\left(\frac{\partial^2 q}{\partial \eta^2} + \frac{\partial^2 q}{\partial \zeta^2}\right) - q|q|^2 + i\gamma(\eta,\zeta)q - i\alpha q|q|^2. \quad (1)$$

Here $\eta, \zeta$ are the normalized transverse coordinates and $\xi$ is the propagation distance; the function $\gamma(\eta,\zeta)$ describes the gain profile; $\alpha$ is the parameter of two-photon absorption. The Eq. (1) can be used to describe the nonlinear response of semiconductor materials where solitons form below half the bandgap and two-photon absorption is the dominating mechanism of optical losses [18]. Suitable shaping of the concentration of amplifying centers may be used to create various gain landscapes in such materials [19].

We consider solitons in a radially symmetric gain landscape described by the function $\gamma = p_i \exp[-(r-r_c)^2/d^2]$, where $p_i$ is the gain parameter, $r = (\eta^2+\zeta^2)^{1/2}$ is the radial variable, $d$ and $r_c$ are the width and the radius of the amplifying ring, respectively. Here we set $r_c = 5.25$ and $d = 1.75$. Soliton solutions of Eq. (1) may exist due to the balance between localized gain and nonlinear losses, and between diffraction and focusing nonlinearity. Spatial localization of the gain ensures stability of the background at $r \to \infty$, a crucial ingredient for the stability of localized solutions. The formation of solitons should be expected around the maxima of $\gamma(\eta,\zeta)$ where nonlinear absorption compensates gain, thus preventing the uncontrolled growth of the soliton amplitude and suppressing collapse. This point must be stressed, since the nonlinearity is cubic (Kerr) and the system is two-dimensional. Since an exact balance between gain and losses is required for soliton formation the soliton shape and transverse extent should depend considerably on the shape of the amplifying domain. The crucial open questions that we address in this Letter are whether such gain landscapes can support dynamically evolving vortex solitons with multiple singularities and whether such solitons may be stable. In order to find rotating vortex solitons we integrate Eq. (1) directly for a variety of inputs $q(\eta,\zeta,\xi=0)$, with single or with multiple phase singularities nested in different locations of a single core. Under proper conditions, input beams emit radiation, experience reshaping upon propagation, and approach rotating vortex soliton states. In our simulations we propagated the beams for large lengths (in some cases up to $\xi \sim 20000$), to confirm stability of the obtained rotating states. We verified that identical rotating solitons can be generated with different inputs, a result that is an indication that the states are attractors with a sufficiently large basin.

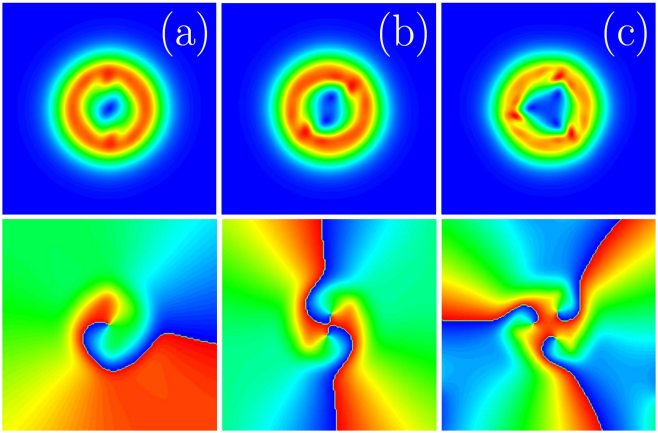

Figure 1. Field modulus (top) and phase distributions (bottom) for rotating vortex solitons having (a) one phase singularity at $p_i = 1.7$, (b) two singularities at $p_i = 2.0$, and (c) three singularities at $p_i = 2.8$. In all cases $\alpha = 2.0$.

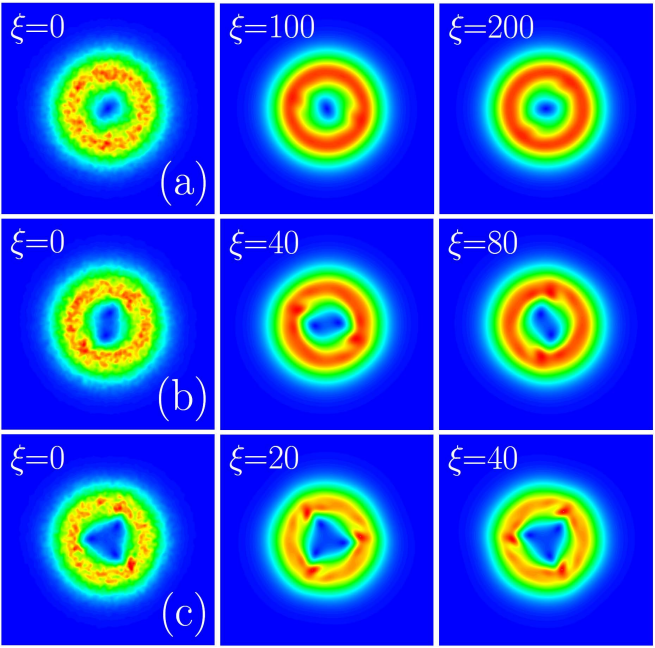

Figure 2. Dynamics of propagation for rotating vortex solitons having (a) one phase singularity at $p_i = 1.7$, (b) two singularities at $p_i = 2.0$, and (c) three singularities at $p_i = 2.8$. In all cases $\alpha = 2.0$ and white noise was added into the input fields.

The central result of this Letter is that the ring-like gain landscapes can support a rich variety of nonconventional rotating vortex solitons. In all such states, the phase singularities are embedded in a single vortex core and perform persistent rotation around the center of the gain ring. Illustrative examples of the simplest rotating structures with one, two, or three phase singularities are presented in Fig. 1. Notice that all such states can be generated with the ring input beams $q \sim r \exp(-r^2) \exp(im\varphi)$ carrying singularities with winding numbers $m = 1, 2, 3$ when their centers are strongly displaced from the center of the amplifying ring at $\eta, \zeta = 0$. The number of singularities in the rotating vortex solitons is not determined exclusively by the charge or number of singularities in the input beam. Actually, rotating vortex with $m > 1$ singularities can be excited even by a shifted vortex with $m = 1$. Notice that in contrast to conservative settings where rotating states emerge from azimuthally perturbed vortex solitons [20], in our dissipative system azimuthal perturbations of radially symmetric states usually tend to decay.

All rotating vortex solitons exhibit azimuthal modulations of the intensity distributions, in spite of the fact that the gain landscape is radially symmetric. The depth of the azimuthal intensity modulation depends on the value of the gain parameter $p_i$. We observed that, as a rule, the azimuthal modulation is most pronounced in the center of the existence domain (in terms of $p_i$) and decreases when approaching the upper boundary of the domain. The decrease of the azimuthal modulation is usually accompanied by a decrease of separation between the phase singularities nested in the vortex core. The phase singularities in the simplest rotating vortex solitons are located symmetrically with respect to the origin at $\eta, \zeta = 0$, but we found that vortex states with an asymmetric arrangement of the singularities may be generated, too. The propagation dynamics of the rotating vortex solitons is illustrated in Fig. 2. Such structures were numerically generated by direct propagation of inputs beams over huge distances, so that a stable solution was excited, and subsequently used them as inputs with added noise. Note that in practice approximate versions of such solutions may be generated by using phase masks. The perturbed solutions were then propagated for one rotation period to ensure that they perform persistent rotation even in the presence of noise. Rotating vortex solitons are characterized by highly nonconventional energy flow distributions $\mathbf{S}_\perp = \mathrm{Im}(q^* \nabla_\perp q)$ [Fig. 4(d)]. They are not circular due to nonzero radial components. There are several maxima of $|\mathbf{S}_\perp|$ in the points where the intensity reaches also a maximum [the hydrodynamic form of Eq. (1) shows that asymmetric internal energy flows affect soliton intensity distributions rendering them asymmetric]. The soliton intensity peaks rotate in the direction of the phase gradient [a counterclockwise rotation can be observed by comparing Fig. 4(d) and dynamics in Fig. 2(c)].

The existence domains of the rotating vortex solitons with two and three phase singularities are presented in Fig. 3. The domains were constructed using slow modifications of the gain parameter until either transformation into radially symmetric vortex states (this scenario is typical for high $p_i$ values) or the development of instabilities is achieved. Therefore, the domains shown in Fig. 3 correspond to the parameter range where the rotating vortex solitons are stable. For a given strength of two-photon absorption, the stable rotating vortex solitons exist inside a limited range of values of the gain coefficient $p_i^{\mathrm{low}} \leq p_i \leq p_i^{\mathrm{upp}}$. The growth of the nonlinear losses shifts this limited existence domain to higher values of gain. Usually, vortex states with three singularities exist at highest gain values, although the existence domains for solitons with two and three singularities may overlap slightly. Similarly, vortices with one singularity exist at lowest gain levels. Therefore, a general conclusion can be made that higher gain values allows generation of more complex rotating vortex solitons with larger number of singularities. Similar existence domains were obtained for other $r_c$ and $d$ values. If $d \ll r_c$ an increase of $d$ results in a shift of the entire existence domains to lower gain values. However, there exist an upper limit ($d \sim 0.8 r_c$) for the width of the amplifying ring (that depends on $p_i$) at which rotating vortex solitons can be generated. The domains of existence of rotating vortices may overlap with domains of existence of radially symmetric states [15], i.e. several stable attractors coexist for the same set of parameters.

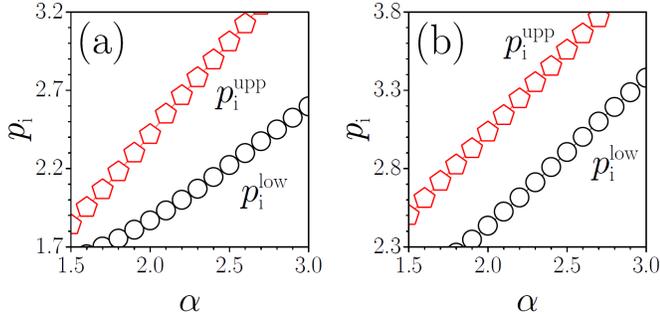

Figure 3. Domains of existence of stable rotating vortex solitons with (a) two and (b) three singularities on the plane $(\alpha, p_i)$.

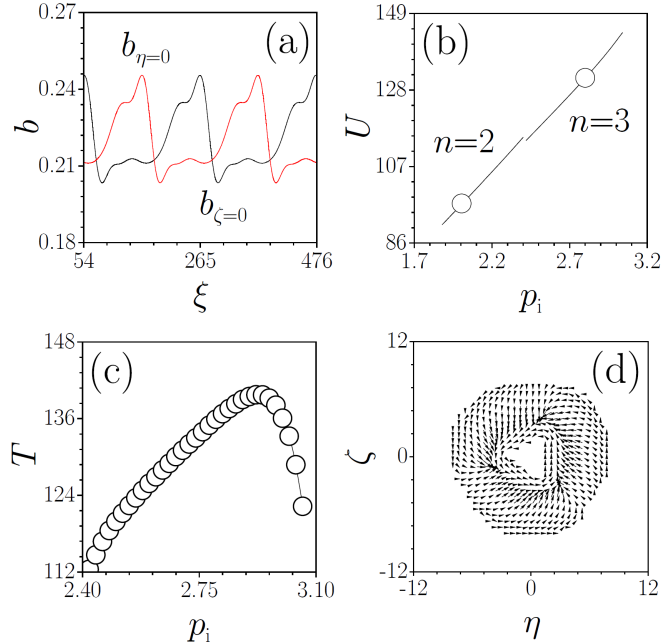

Figure 4. (a) Instantaneous propagation constant value defined in different points versus propagation distance for rotating vortex with one singularity at $p_i = 1.7$, $\alpha = 2.0$. (b) Power versus $p_i$ for vortex solitons with two or three singularities at $\alpha = 2.0$. Points correspond to solitons shown in Figs. 1(b) and 1(c). (c) The period of rotation for vortex with three singularities versus $p_i$ at $\alpha = 2.0$. (d) Map of internal energy flows in vortex soliton from Fig. 1(c).

In contrast to the conventional nonrotating vortex solitons characterized by a single propagation constant value, the local propagation constant of the rotating vortex soliton, defined as $b(\xi) = \mathrm{Arg}[q(\xi + d\xi)/q(\xi)]/d\xi$, is not identical in all points in the transverse plane. This propagation constant evolves periodically with distance $\xi$ that may be attributed to the rotation of the entire azimuthally modulated complex. Thus, Fig. 4(a) shows the propagation constant values $b_{\eta=0}$ and $b_{\zeta=0}$ defined along the corresponding lines as functions of $\xi$. The shift between these two periodic dependencies is readily apparent. The power $U = \int\int_{-\infty}^{\infty} |q|^2 d\eta d\zeta$ is a monotonically increasing function of the gain parameter irrespectively of the number of singularities in the rotating vortex [Fig. 4(b)]. As expected on intuitive grounds, more complicated vortex structures have higher powers.

We also studied the dependence of the rotation period $T$ of the vortex states on the gain parameter [Fig. 4(c)]. In the vicinity of the lower border of the existence domain $p_i \to p_i^{low}$ the rotation period grows with $p_i$ monotonically, but when $p_i \to p_i^{upp}$ the rotation period rapidly decreases. A decrease in the period of rotation favors the tendency for singularities to fuse into one central singularity with a higher winding number at sufficiently high gain levels (in this case the rotating vortex transforms into the radially symmetric one). The dependencies $T(p_i)$ shown in Fig. 4(c) are typical for all vortex solitons discussed in this work.

### References

1) A. S. Desyatnikov, Y. S. Kivshar, and L. Torner, Prog. Opt. **47**, 291 (2005).
2) N. Akhmediev and A. Ankiewicz, eds., *Dissipative Solitons: From Optics to Biology and Medicine*, Vol. 751 of Lecture Notes in Physics (Springer, Berlin, 2008).
3) N. N. Rosanov, *Spatial hysteresis and optical patterns*, (Springer, Berlin, 2010).
4) N. N. Rosanov, S. V. Fedorov, and A. N. Shatsev, Phys. Rev. Lett. **95**, 053903 (2005).
5) J. Keeling and N. G. Berloff, Phys. Rev. Lett. **100**, 250401 (2008).
6) A. Alexandrescu and V. M. Perez-Garcia, Phys. Rev. A **73**, 053610 (2006).
7) L. C. Crasovan, B. A. Malomed, and D. Mihalache, Phys. Rev. E **63**, 016605 (2001).
8) J. M. Soto-Crespo, N. Akhmediev, C. Mejia-Cortes, and N. Devine, Opt. Express **17**, 4236 (2009).
9) V. Skarka, N. B. Aleksic, H. Leblond, B. A. Malomed, and D. Mihalache, Phys. Rev. Lett. **105**, 213901 (2010).
10) W. C. Mak, B. A. Malomed, and P. L. Chu, Phys. Rev. E **67**, 026608 (2003).
11) C.-K. Lam, B. A. Malomed, K. W. Chow, and P. K. A. Wai, Eur. Phys. J. Special Topics **173**, 233 (2009).
12) Y. V. Kartashov, V. V. Konotop, V. A. Vysloukh and L. Torner, Opt. Lett. **35**, 1638 (2010).
13) Y. V. Kartashov, V. V. Konotop, V. A. Vysloukh, and L. Torner, Opt. Lett. **35**, 3177 (2010).
14) Y. V. Bludov, and V. V. Konotop, Phys. Rev. A **81**, 013625 (2010).
15) V. E. Lobanov, Y. V. Kartashov, V. A. Vysloukh, and L. Torner, Opt. Lett. **36**, 85 (2011).
16) F. Ye, Y. V. Kartashov, B. Hu, and L. Torner, Opt. Lett. **35**, 628 (2010).
17) T. J. Alexander, A. S. Desyatnikov, and Y. S. Kivshar, Opt. Lett. **32**, 1293 (2007).
18) O. Katz, Y. Lahini, and Y. Silberberg, Opt. Lett. **33**, 2830 (2008).
19) M. J. Connelly, *Semiconductor Optical Amplifiers* (Springer, Berlin, 2002).
20) A. S. Desyatnikov, A. A. Sukhorukov, and Y. S. Kivshar, Phys. Rev. Lett. **95**, 203904 (2005).